\documentclass[12pt]{article}

\usepackage{scicite}

\usepackage{times}

\newenvironment{sciabstract}{%
\begin{quote} \bf}
{\end{quote}}

\usepackage{graphicx}
\usepackage{bm}
\usepackage[matrix,frame,arrow]{xy}
\usepackage[applemac]{inputenc}
\usepackage[T1]{fontenc}
\usepackage{lmodern}
\usepackage[english]{babel}
\usepackage{amsmath}
\usepackage{ae}
\usepackage{units}
\usepackage{amssymb}
\usepackage{color}
\usepackage{bbm}
\usepackage{url}
\usepackage{fixmath}
\usepackage{booktabs}

\usepackage[colorlinks]{hyperref}

\newcommand{\be}{\begin{equation}}
\newcommand{\ee}{\end{equation}}
\newcommand{\bea}{\begin{eqnarray}}
\newcommand{\eea}{\end{eqnarray}}

\newcommand{\sz}{\sigma_z}
\newcommand{\sm}{\sigma_-}
\renewcommand{\sp}{\sigma_+}

\newcommand{\ket}[1]{|#1\rangle}

\newcommand{\expec}[1]{\left\langle #1 \right\rangle}

\newcommand{\abs}[1]{\left|#1\right|}

\renewcommand{\eqref}[1]{\mbox{Eq.~(\ref{#1})}}

\newcommand{\tabref}[1]{\mbox{Table~\ref{#1}}}

\newcounter{lastnote}

\title{Probing the quantum vacuum with an artificial atom in front of a mirror} 

\author
{I.-C. Hoi$^{1,3}$, A. F. Kockum$^{1}$, L. Tornberg$^{1}$,
A. Pourkabirian$^{1}$, G. Johansson$^{1}$\\
P. Delsing$^{1,\ast}$ \& C. M. Wilson $^{2,\ast}$\\
\\
\normalsize{$^{1}$Department of Microtechnology and Nanoscience (MC2),}\\
\normalsize{Chalmers University of Technology, SE-412 96 Gothenburg, Sweden}\\
\normalsize{$^{2}$Institute for Quantum Computing and Electrical and Computer Engineering Department,}\\
\normalsize{University of Waterloo, Waterloo, Canada}\\
\normalsize{$^{3}$Department of Physics, University of California,}\\
\normalsize{Santa Barbara, California 93106, USA}\\
\\
\normalsize{$^\ast$To whom correspondence should be addressed}\\
\normalsize{E-mail: chris.wilson@uwaterloo.ca; per.delsing@chalmers.se}\\
}

\date{}

\begin{document} 


\baselineskip24pt


\maketitle


\begin{sciabstract}
Quantum fluctuations of the vacuum are both a surprising and fundamental phenomenon of nature. Understood as virtual photons flitting in and out of existence, they still have a very real impact, \emph{e.g.},  in the Casimir effects and the lifetimes of atoms. Engineering vacuum fluctuations is therefore becoming increasingly important to emerging technologies.  Here, we shape vacuum fluctuations using a "mirror" , creating regions in space where they are suppressed.  As we then effectively move an artificial atom in and out of these regions, measuring the atomic lifetime tells us the strength of the fluctuations. The weakest fluctuation strength we observe is 0.02 quanta, a factor of 50 below what would be expected without the mirror, demonstrating that we can hide the atom from the vacuum. 

%
\end{sciabstract}

\section*{}
From the earliest days of exploration of quantum electrodynamics, it was thought that quantum fluctuations of the vacuum could have important physical effects, for instance, determining the lifetimes of excited states of atoms \cite{Dirac}, giving rise to the Lamb shift \cite{Bethe, Lamb}, and modifying the gyromagnetic ratio of the electron \cite{Schwinger,Welton}. This invocation of the vacuum to explain measurable physical effects was controversial. In the intervening years, however, the idea that the vacuum itself is physical gained increasing credence with a growing number of striking vacuum phenomena predicted such as Hawking radiation \cite{Hawking}, the Unruh effect \cite{Unruh} and the Casimir effects \cite{StaticCasimir,DynamicalCasimir}.  In recent years, these vacuum effects have even started to have technological impacts, contributing to stiction in nanomechanics \cite{StictionCasimir} and limiting the coherence times of superconducting qubits \cite{AHouck,Wilson}. This has led to an increasing interest in engineering the vacuum.  In this work, we demonstrate engineering of the mode structure of the quantum vacuum. We show that we can shape the modes of the vacuum itself using a "mirror". We use a superconducting qubit as a sensitive probe of the vacuum modes. Further, we show that we can hide the qubit from these quantum vacuum fluctuations using this technique.

The effect of a mirror on the radiative decay of natural atoms has been studied previously\cite{Eschner,Dorner,Bushev,Dubin,Glaetzle}. While achieving impressive results, the work was limited by the small solid angle of the atomic radiation that could be made to interact with the mirror.  In our work, this problem is solved by strongly coupling our artificial atom, the qubit, to a one-dimensional superconducting waveguide that collects $>99\%$ of the radiation from the atom.  In addition, we eliminate any motional noise, as the qubit and mirror are fixed in place. This allows us to observe a modulation of the excited-state lifetime by a factor of 9.8. In contrast to the Purcell effect \cite{Purcell, Kleppner, Martini, Lee}, where a cavity is used to modify the lifetime, in this work we have a fully open system with a continuous spectrum of modes.  Recent theoretical work has also suggested other novel ways to suppress the effects of vacuum fluctuations on an artificial atom\cite{GiantAtom}.



\begin{figure}[h!]
\centering
 \includegraphics[width=0.5\columnwidth]{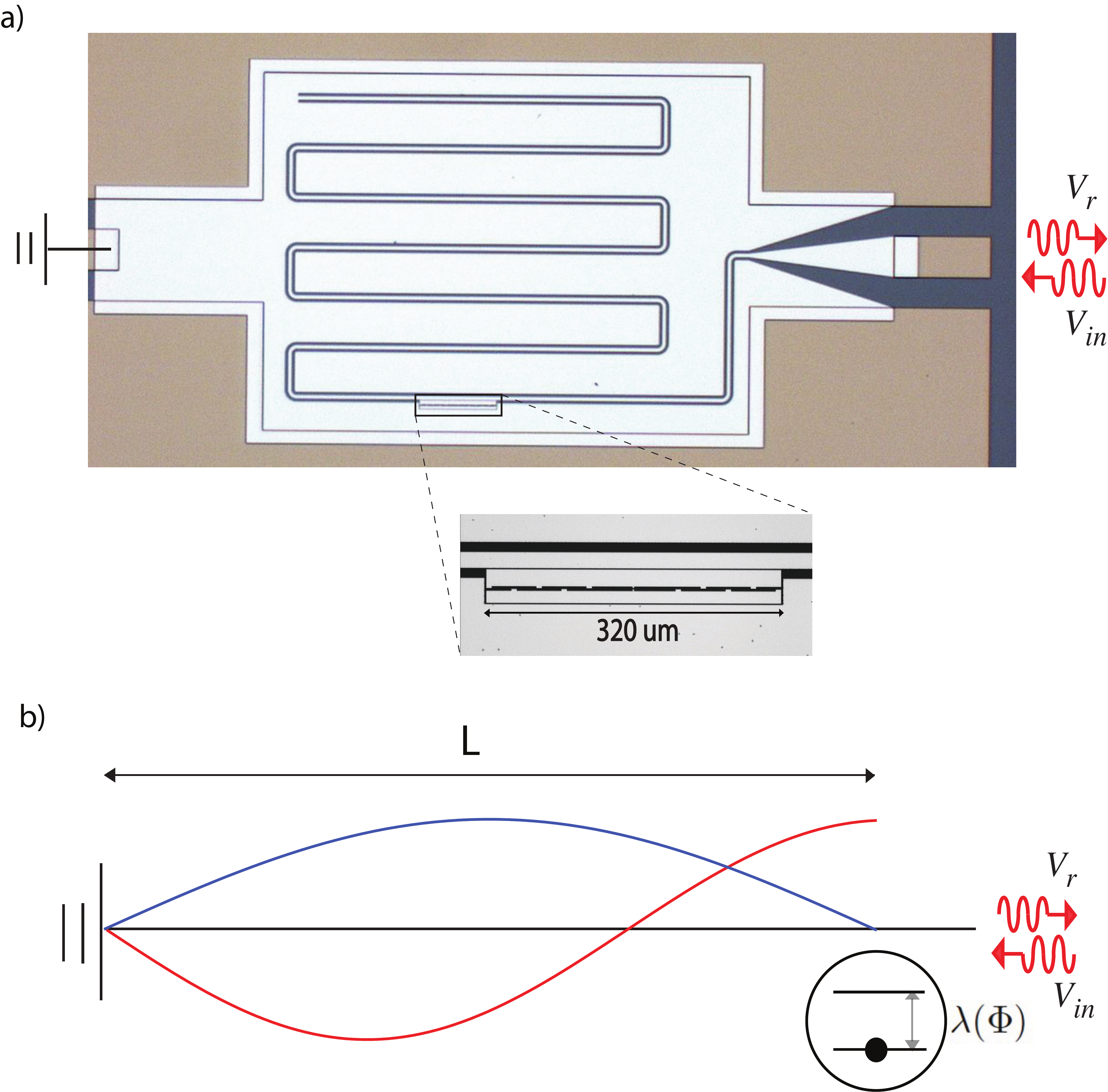}
\caption{An artificial atom in front of a mirror. {\bf a}) A micrograph of the atom-mirror system, a superconducting transmon embedded at a distance $L$ from the end of a 1D transmission line. (zoom in) The transmon. The atom size is small compared to the wavelength of the microwave field. We characterize the system by sending in a coherent probe field, $V_{in}$, at $\omega_p \approx 5$\,GHz and measuring the reflected field, $V_{r}$.  Measurements are done at $T=50$\,mK, where thermal excitations of the field are negligible.
{\bf b}) Cartoon of the atom-mirror system. The blue and red curves show the mode structure of the voltage along the transmission line at the atom frequency for $L=\lambda/2$ and $L=3\lambda/4$, respectively. By tuning $\lambda$ of the two-level atom via an external magnetic flux, $\Phi$, the coupling between the field and the atom can be turned off when the atom sits at a node of the resonant EM field (blue). The atom is maximally coupled at the antinode (red).
\label{FigExpSetup}}
\end{figure}

In this article, following the new paradigm of waveguide quantum electrodynamics (wQED) \cite{Astafiev1,Abdumalikov, Hoi, Hoi2, Hoi3, Hoi1,Arjan, Huaixiu, Chang, Shen2, Kevin}, we study an artificial atom, a superconducting transmon \cite{Koch}, embedded at a distance $L$ from the end of a $Z_0=50$\,$\Omega$ transmission line, where the center conductor is short-circuited to the ground plane. This imposes a reflecting boundary condition on the electromagnetic (EM) field in the line, creating the equivalent of a mirror. (A micrograph of the device is shown in Fig.~\ref{FigExpSetup}a.) In particular, interference between an incoming field and the field reflected by the mirror creates a standing-wave pattern, with a voltage node at the mirror plane and a voltage amplitude that varies periodically along the line (See Fig.~\ref{FigExpSetup}b). Crucially, quantum electrodynamics tells us that this mode structure is imposed not only on any classical field in the line, but also on the vacuum fluctuations of the field.  While the structure of the vacuum fluctuations cannot be directly measured with a classical probe, like a voltmeter, they can be measured by observing the effect of the vacuum fluctuations on a quantum probe, such as an atom or qubit.  The decay rate of an excited state, $\ket{1}$, with a transition frequency $\omega_{a}$ to the ground state, $\ket{0}$, is proportional to the strength (spectral density) of EM fluctuations near the frequency $\omega_{a}$ that are present in the atom's environment.  If the atom is in an environment at a temperature $T \ll \hbar\omega_{a}/k_B$, the excited state lifetime of the atom will be limited by vacuum fluctuations because (classical) thermal fluctuations of the field are exponentially suppressed at these temperatures.  Therefore, measuring the lifetime of the atom, which can be done through conventional spectroscopy, probes the local strength of vacuum fluctuations at the transition frequency.  In effect, the atom acts as a quantum spectrum analyzer.

To probe the spatial structure of the modes, we need to change the effective distance between the atom and mirror. While it is difficult to change the physical distance, $L$,  \textit{in situ}, the relevant quantity is in fact the \textit{normalized} distance, $L/\lambda$, where  $\lambda$ is the transition wavelength of the atom.  We can easily change $\lambda$ by tuning $\omega_a$ with an external magnetic flux perpendicular to the transmon. As illustrated in Fig.~\ref{FigExpSetup}b, tuning $\lambda$ allows us to effectively move the qubit from a node to an antinode of the resonant vacuum fluctuations.  By measuring the qubit lifetime as a function of frequency, we can therefore map out the frequency-dependent spatial structure of the vacuum.


In detail, the transition wavelength of the transmon can be expressed as \cite{Koch}
\be
\lambda \left( \Phi \right)=2\pi v/\omega_{a} \left( \Phi \right) \simeq h v / \left(\sqrt{8 E_C E_J(\Phi)}-E_C \right),
\label{EqLambda}
\ee
where $h$ is Planck's constant, $v=c/\sqrt{\epsilon}$ is the velocity of the wave propagating along the transmission line, $\epsilon$ is the effective dielectric constant of the transmission line, and $c$ is the velocity of light in vacuum. $E_C$ and $E_J(\Phi)$ are the charging and Josephson energies of the transmon, respectively, and $E_J(\Phi)=E_{J,0} \abs{\cos(\pi\Phi/\Phi_0)}$, where $E_{J,0}$ is the maximum Josephson energy, $\Phi$ is the magnetic flux and $\Phi_0=h/(2e)$ is the flux quantum. 

We characterize the system spectroscopically by sending a coherent microwave field toward the transmon and measuring the reflection coefficient, $r_p=\expec{V_r}/\expec{V_{in}}$, where $\expec{V_r}$ ($\expec{V_{in}}$) is the time-averaged reflected (incident) field. Note that $r_p$ is a phase-sensitive average and, therefore, only captures the coherently scattered signal. As demonstrated in previous experiments, all the fields are reflected either coherently or incoherently and losses are neglected in the rest of the paper \cite{Astafiev1,Hoi1}. 

Consider the situation depicted in Fig.~\ref{FigExpSetup}. The coherent input $V_{in}$ interacts with the atom and then continues moving to the left. The scattered field from the atom, proportional to $\expec{\sm}$ (the expectation value of the atomic lowering operator), is equally divided between left- and right-moving states. $V_{in}$ and the left-moving field from the atom are then reflected at the mirror and return to interact with the atom once more. Since the roundtrip time is small compared to the timescale of the atomic evolution,  we only need to take into account the phase factor
\be
\theta(\Phi)=2\times[2\pi L/\lambda(\Phi)]+\pi,
\label{EqTheta} 
\ee
which the field acquires during the roundtrip. Here, the added $\pi$ phase shift is due to the reflection at the mirror. Summing up all the fields to get the output, we arrive at the reflection coefficient \cite{Kazuki,Hoi3}
\be
r_p=-[1+2\Gamma_{1}\left\langle \sm \right\rangle/\Omega_p]e^{i4\pi L/\lambda},
\label{EqReflection}
\ee
where $1/\Gamma_{1}$ is the excited-state lifetime of the atom which is dominated by the coupling to the transmission line via the coupling capacitor $C_c$. $\Omega_p$ is the Rabi frequency, which is proportional to the probe amplitude $V_{in}$. The phase term $e^{i4\pi L/\lambda}$ in \eqref{EqReflection} does not affect the dynamics and is removed. However, the phase factor $\theta$ is still present in the definitions of $\Gamma_{1}$ and $\Omega_p$. The dynamics of the scattered field is governed by $\expec{\sm}$, which is found by solving the Bloch equations
\bea
\partial_t \expec{\sigma_\pm} &=& \left(\pm i\delta\omega_p - \gamma \right)\expec{\sigma_\pm} + \Omega_p \expec{\sz}/2, 
\label{EqBlochPM}
\\
\partial_t \expec{\sz} &=& -\Gamma_{1} \left( 1+ \expec{\sz} \right) - \Omega_p \left(\expec{\sp} + \expec{\sm}\right),
\label{EqBlochZ}
\eea
where $\sp$ is the atomic raising operator, $\sz$ is the third Pauli spin operator, $\gamma=\Gamma_{1}/2+\Gamma_\phi$ is the decoherence rate, with $\Gamma_\phi$ being the pure dephasing rate, and $\delta\omega_p=\omega_{a}-\omega_p$ is the detuning between $\omega_{a}$ and the probe frequency, $\omega_p$.  

The inverse lifetime, $\Gamma_{1}$, is the quantity of greatest interest to us, as it is proportional to the strength of vacuum fluctuations. The dephasing rate, $\Gamma_{\phi}$, is instead related to low-frequency fluctuations in the environment \cite{Makhlin}.

\begin{figure*}[ht!]
 \includegraphics[width = \columnwidth]{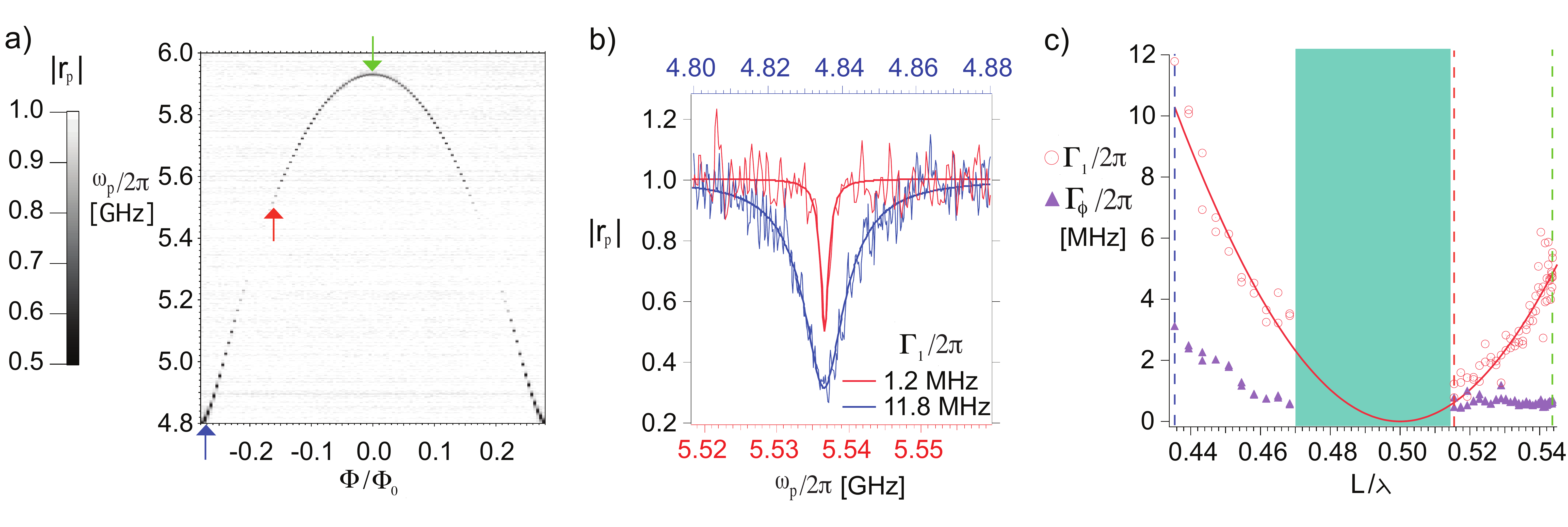}
\caption{Spectroscopic measurements of the excited-state lifetime. {\bf a}) The reflection coefficient $|r_p|$ as a function of $\omega_p$ and $\Phi$ for a weak probe ($\Omega_p\ll\gamma$). Because the atomic linewidth is much less than the tuning range, the qubit response appears as a narrow black line against the white background, which corresponds to $|r_p|\approx1$. As we tune $\Phi$, $\lambda$ varies according to \eqref{EqLambda}. When $L\approx\lambda/2$, which corresponds to 5.4\,GHz, the qubit sits at the node of the field and, therefore, is hidden from the probe and no signal is observed. {\bf b}) $|r_p|$ as a function of $\omega_p$ at two values of $\Phi$, indicated by the blue and red arrows in (a). The solid curves are theoretical fits using \eqref{EqReflectionWeakProbe}, from which we extract $\Gamma_{1}$, $\gamma$ and $\omega_{a}$, where $\gamma=\Gamma_{1}/2+\Gamma_{\phi}$.  At the low temperatures of our experiment, the inverse lifetime $\Gamma_{1}$ is proportional to the strength of the vacuum fluctuations. We see $\Gamma_{1}$ changing by a factor of 9.8 between these two flux biases, indicating a large modulation in the amplitude of vacuum fluctuations, which is due to the frequency dependence of the spatial mode structure. {\bf c}) For each flux bias in (a), similar to the procedure in (b), we extract $\Gamma_{1}(\Phi)$ and $\Gamma_\phi(\Phi)$, denoted by the red and purple markers, respectively. We plot these rates as a function of the normalized distance, $L/\lambda(\Phi)$. The solid red curve is theory based on \eqref{EqGammaPhiDependence}. The red and blue dashed lines indicate the two cases displayed in (b). The green arrow in (a) and the green dashed line in (c) indicate the flux bias point for Fig.~\ref{FigCoherentScattering}. The shaded blue region indicates where the response from the atom is too weak to measure; this is where the atom is hidden from the vacuum fluctuations. 
\label{FigTunableCoupling}}
\end{figure*}

In Fig.~\ref{FigTunableCoupling}a, we plot $\abs{r_p}$ as a function of $\omega_{a}$ and $\Phi$ for a weak probe, \emph{i.e.}, $\Omega_p\ll\gamma$. We see a response from the atom for $\omega_{a}/2\pi$ ranging from $4.8$\,GHz to $5.93$\,GHz. The atomic response becomes weaker and weaker when $\omega_{a}/2\pi$ approaches the region around $5.4$\,GHz, and eventually the response vanishes, because at this frequency the qubit sits at the node of the probe voltage. In this way, we see that we can hide the atom from the classical probe field even though it sits fully exposed in an open transmission line.  This phenomenon can be described as an interference between the atom and its mirror image.

To understand the effects of vacuum fluctuations, we must look in more detail at the spectroscopic line shape of the atom. In Fig.~\ref{FigTunableCoupling}b, we plot $\abs{r_p}$ as a function of $\omega_p$ for two flux biases. These data are line cuts in Fig.~\ref{FigTunableCoupling}a, indicated by the blue and red arrows. In the steady state, where $\partial_t\expec{\sigma_i}=0$, $i=\pm,z$, Eqs.~(\ref{EqReflection})-(\ref{EqBlochZ}) give for a weak probe
\be
r_p = -1+\frac{\Gamma_{1}}{\gamma + i \delta\omega_p}.
\label{EqReflectionWeakProbe}
\ee
The solid curves in Fig.~\ref{FigTunableCoupling}b are fits to the data using \eqref{EqReflectionWeakProbe}. The width of the peak gives us $\gamma$ directly. The depth of the dip gives the ratio $\Gamma_{1}/\gamma$, allowing us to extract  $\Gamma_{1}$ directly. We note that $\Gamma_{1}$ changes by a factor of $9.8$ between the two different flux bias points, indicating a large modulation in the strength of vacuum fluctuations. In the region around $5.4$\,GHz, $\Gamma_{1}$ is even smaller and approaches zero. However, since the coupling is so small, we can no longer measure it. In this region, as expected from \eqref{EqReflectionWeakProbe}, $|r_p|\simeq1$ and the atom, in concert with its mirror image, hides from the field.

In Fig.~\ref{FigTunableCoupling}c, we use \eqref{EqReflectionWeakProbe} to extract $\Gamma_{1}(\Phi)$, $\Gamma_\phi(\Phi)$, and $\omega_{a}(\Phi)$ for each flux bias in Fig.~\ref{FigTunableCoupling}a. In the shaded blue region around $\omega_{a}/2\pi=5.4$\,GHz, the qubit is hidden and we cannot extract any data. The inverse lifetime varies as a function of $\Phi$ according to \cite{Kazuki}
\be
\Gamma_{1}(\Phi)=2\Gamma_{1,b}\cos^2[\theta(\Phi)/2],
\label{EqGammaPhiDependence}
\ee
where $\Gamma_{1,b}$ is the inverse of the bare atomic lifetime. This shows how we can tune the inverse lifetime between $0$ and $2\Gamma_{1,b}$ (corresponding to $\lambda = 2L$ and $\lambda=4L$, respectively) by tuning the flux. The factor of two comes from the enhancement of the vacuum fluctuations, due to constructive interference between the atom and its mirror image, which is not present in the absence of the mirror. Recently, a similar interference effect has also been observed with two artificial atoms in an open line \cite{Arjan}. 

In \tabref{Params}, we summarize the parameters extracted from the data in Fig.~\ref{FigTunableCoupling}. The value of $\Gamma_{1,b}$ is consistent with what we measured in a separate experiment with a very similar transmon at the end of an open-circuited transmission line (antinode) \cite{Hoi3}, where we extracted $\Gamma_{1} = 63$\,MHz $\sim 2\Gamma_{1,b}$.\\

\begin{table}[ht!]
\centering
\begin{tabular}{| c | c | c | c | c | c |}
\hline
 $E_{J,0}/h$ [GHz] & $E_C/h$ [GHz] & $\omega_a(0)/2\pi$ [GHz] & $\Gamma_{1,b}/2\pi$ [MHz] & $\epsilon$ & $L$ [mm] \\
\hline
 $13.1$ & $0.38$ & $5.93$ & $33$ & $6.25$ & $11$\\
\hline
\end{tabular}
\caption{\label{Params} Parameters of the device.}
\end{table}

\begin{figure}[ht!]
 \centering
 \includegraphics[width=0.5\columnwidth]{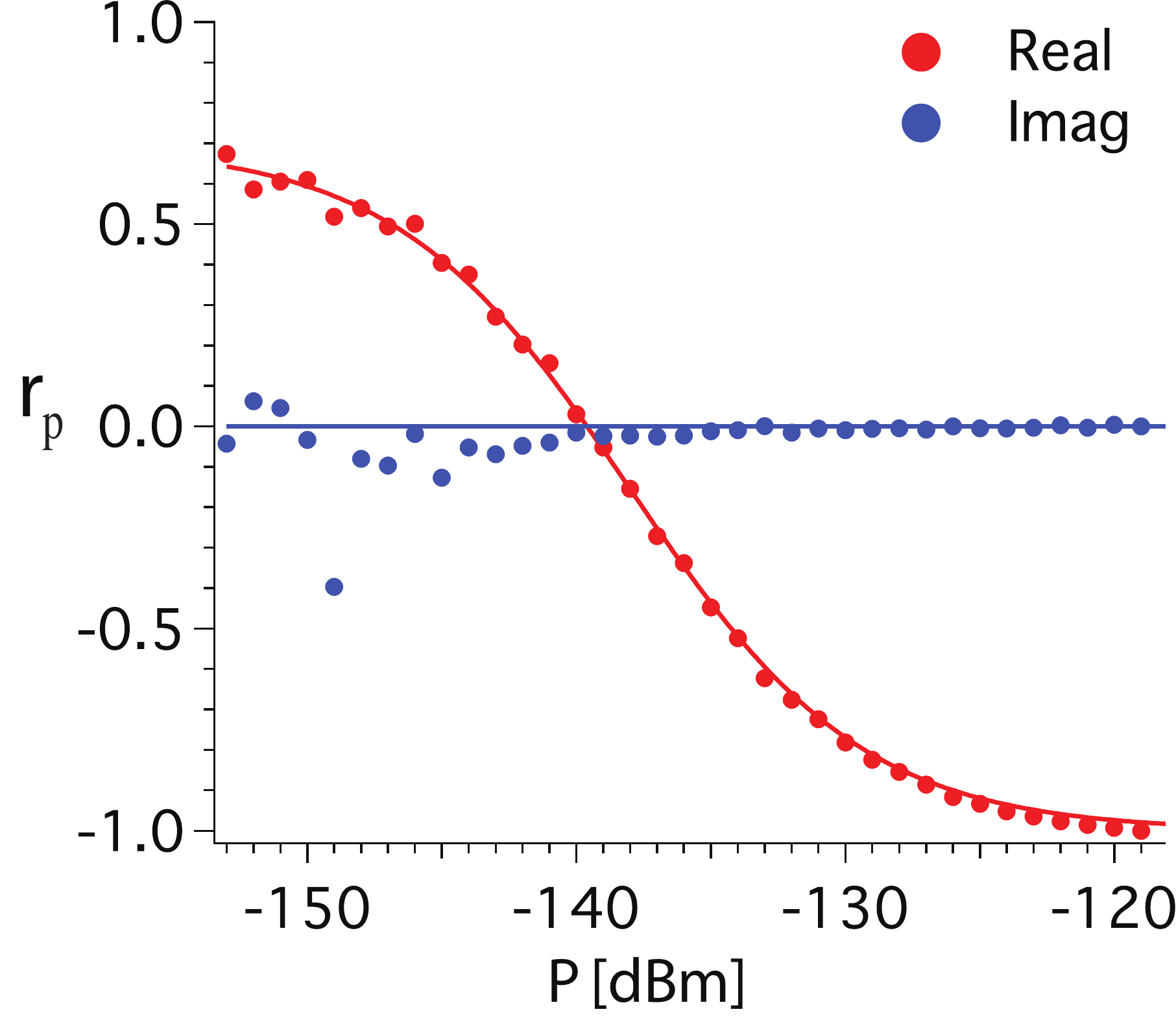}
\caption{Calibrating the atom-field coupling. We measure the nonlinear scattering properties of the atom at $\Phi\simeq0$. The plot shows $r_p$ as a function of resonant incident power, $P$. The real and imaginary response are shown in red and blue, respectively. The markers are experimental data and the solid curves are a theoretical fit based on \eqref{EqReflectionOnResonance}. We use the parameters extracted independently in Fig.~2c (green arrow), leaving as the one free parameter the atom-field coupling, $k$, defined through the relation $\Omega_p = k\sqrt{P}$. At weak incident power, where $\Omega_p\ll\gamma$, the atom reflects mostly coherently. As the incident power increases, $\abs{r_p}$ decreases down to zero and then increases again. From \eqref{EqReflectionOnResonance}, we see that the zero occurs at $\Omega_p=\sqrt{\Gamma_{1}^2-\Gamma_{1}\gamma}$, where the atom scatters all of the field incoherently. At high power, $\Omega_p\gg\gamma$, the atom is saturated by the incident field. Most of the field is simply reflected by the mirror, resulting in $r_p$ approaching $-1$. \label{FigCoherentScattering}}
\end{figure}

The inverse lifetime, $\Gamma_{1}$, is proportional to the strength of EM fluctuations that are present in the atom's environment near the frequency $\omega_{a}$.  The strength is quantified in terms of the spectral density of the fluctuations, $S(\omega_a)$.  We can relate $\Gamma_1$ to $S(\omega_a)$ through the atom-field coupling constant, $k$, using the relation $\Gamma_{1}=k^2S(\omega_a)$  \cite{Wendin}.  To extract $S(\omega_a)$ experimentally, we must therefore measure $k$ in our system. We can do this using the nonlinear scattering properties of our artificial atom.  In Fig.~\ref{FigCoherentScattering}, we plot $\abs{r_p}$ as a function of the incident resonant power for the flux bias $\Phi\simeq0$ (indicated by green arrows in Fig.~\ref{FigTunableCoupling}a). This nonlinear power dependence allows us to extract $k$. In particular, for a resonant field ($\delta\omega_p = 0$), Eqs.~(\ref{EqReflection})-(\ref{EqBlochZ}) give
\be
r_p = -1+\frac{\Gamma_{1}^2}{\Gamma_{1}\gamma+\Omega_p^2}.
\label{EqReflectionOnResonance}
\ee
For low power ($\Omega_p\ll\gamma$), we expect $r_p$ to approach the asymptotic (positive) value determined by the ratio $\Gamma_{\phi}/\Gamma_1$ (see above). As the power increases, $r_p$ decreases, due to increased incoherent scattering, until the coherently reflected signal is zero \cite{Hoi1}.  At this point, all of the incoming probe is absorbed by the atom and reemitted spontaneously with a random phase.  Beyond this point, $r_p$ becomes negative and its magnitude increases again as the atom saturates and cannot absorb all of the incoming photons. Using the extracted values for $\Gamma_{1}$ and $\Gamma_\phi$ at the green dashed line in Fig.~\ref{FigTunableCoupling}c, \eqref{EqReflectionOnResonance} gives the solid curves in Fig.~\ref{FigCoherentScattering}.  Fitting these curves allows us to calibrate the atom-field coupling constant $k$ through the relation $\Omega_p=k\sqrt{P}$.  Through this procedure, we extract $k_e\simeq6.1\times10^{15}$\,Hz/$\sqrt{\text{W}}$, where the subscript ``$e$" denotes the experimental value. However, the absolute value of the incident power $P$ at the sample has an uncertainty of a few dB, contributing a significant uncertainty to this value. 

To reduce the uncertainty, we can alternatively calculate $k$ from its definition in terms of circuit parameters\cite{Peropadre}, $k=e\beta\sqrt{Z_0}(E_J/2E_C)^{1/4}/\hbar$. $E_J$ and $E_C$ are directly measured through the spectroscopic data in Fig. 2 (see table 1).  $Z_0 = 50$ $\Omega$ is well determined by the geometry of the transmission line. We then use Microwave Office, a commercial EM simulation software package, to evaluate the coupling coefficient $\beta=C_c/C_{\Sigma}\simeq0.4$. Note the we use the simulation to evaluate only the capacitance \textit{ratio} which is more accurate than simulating absolute capacitances.  Together with parameters in table 1, this gives $k_s \simeq8.8\times10^{15}$\,Hz/$\sqrt{\text{W}}$, where the subscript ``$s$" denotes the simulated value. The ratio of $k_s$ and $k_e$ is 1.4, which is reasonable for cryogenic microwave experiments. We use the average between $k_s$ and $k_e$ and use the difference as the systematic error bar. This gives $k_m=(7.45\pm1.35)\times10^{15}$\,Hz/$\sqrt{\text{W}}$, where the subscript ``$m$" denotes the mean value.

\begin{figure}
 \centering
 \includegraphics[width=\columnwidth]{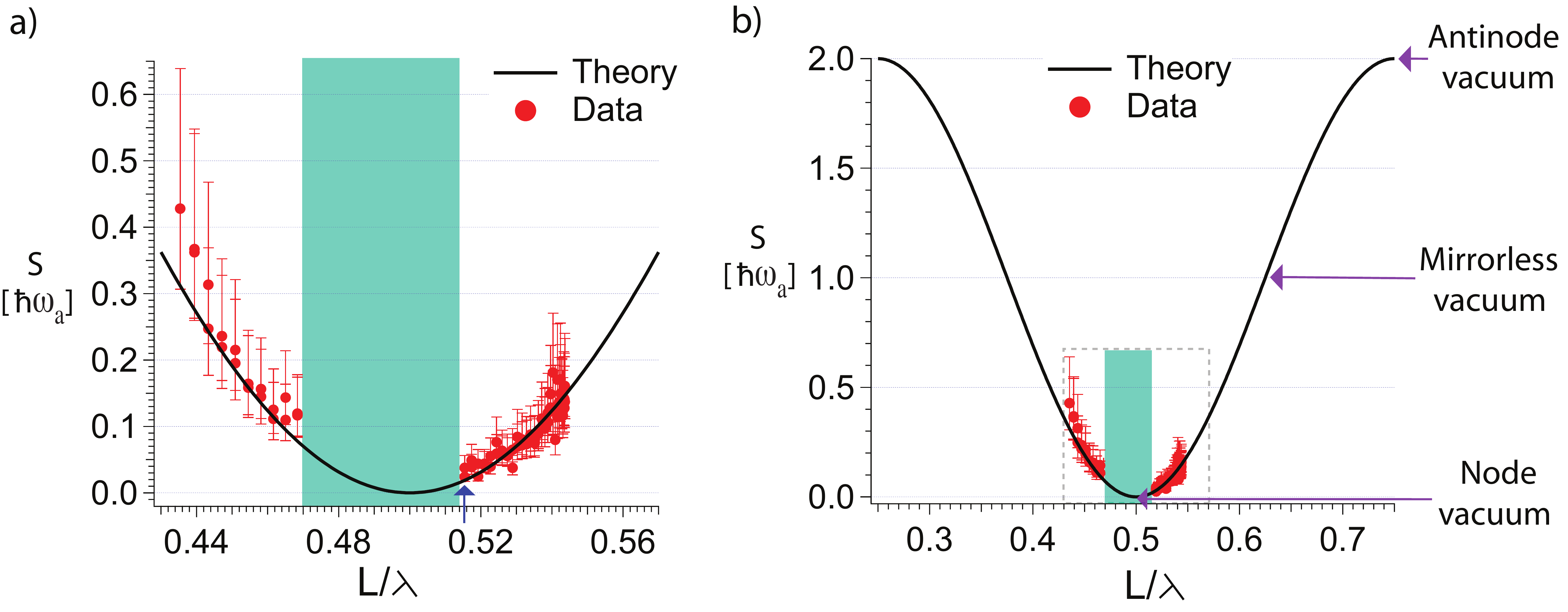}
\caption{The measured spectral density of the vacuum fluctuations $S(\omega_a)$ (red markers) as a function of $L/\lambda$. (The shaded blue region is the same as figure 2c.) $S(\omega_a)$ is displayed in units of number of quanta by normalizing it to $\hbar\omega_{a}$.  In the absence of the mirror, we expect $S(\omega_a)=1$ with half a quanta coming from each side of the transmission line. The error bars indicate the uncertainty of $S$ arising from uncertainty in $\beta$ and the overall attenuation. The solid black curve is the theoretical prediction, without adjustable parameters, according to Eqs.~(\ref{EqSpectrum}) normalized to $\hbar\omega_{a}$. We see that the prediction is inside the error bars, indicating a good agreement between experiment and theory.  The lower limit of the observed spectral density is $S = 0.02$ quanta, indicated by the blue arrow in (a), which is a factor of 50 below the value expected without the mirror. In (b), S oscillates between $2\hbar\omega_a$ and $0$ as a function of $L/\lambda$. The purple arrows in (b) indicate the vacuum fluctuation at $L/\lambda=0.75$ (antinode), $L/\lambda=0.625$ (free space) and $L/\lambda=0.5$ (node), respectively. (a) is the magnification of the dashed square region of (b). 
\label{FigSpectrum}}
\end{figure}

Using $k_m$ and the extracted values of $\Gamma_{1}$ in Fig. 2c, we plot the measured values of $S$ as a function of $L/\lambda$ in Fig.~\ref{FigSpectrum}.  We plot $S(\omega_a)$ in units of number of quanta by normalizing it to $\hbar\omega_{a}$. For an atom in an open line with no mirror, we expect $S = 1$ quanta. The error bars indicate the uncertainty in $S$ arising from the uncertainty in $k_m$. From theory \cite{Clerk}, we expect the spectral density to be

\be
S(\omega_a)=2\hbar\omega_a\cos^2[\theta(\Phi)/2],
\label{EqSpectrum}
\ee
which is shown by the solid black curve in Fig.~\ref{FigSpectrum}. Fig.~\ref{FigSpectrum}a is the magnification of the dashed square region of Fig.~\ref{FigSpectrum}b. In Fig.~\ref{FigSpectrum}b, we show a wider range of normalized distance. We see that the vacuum fluctuations at $L/\lambda=0.75$ (antinode), $L/\lambda=0.625$ (free space), and $L/\lambda=0.5$ (node) are $2\hbar\omega_a$, $\hbar\omega_a$ and $0$, respectively, as indicated by the purple arrows. We see that the black curve falls inside the error bars, indicating a good agreement between experiment and theory and demonstrating that the atomic lifetime is dominated by the spatially-engineered vacuum fluctuations. 
 
 
In conclusion, we have shown that we can shape the modes of the quantum vacuum using a mirror. We have used an artificial atom placed in front of the mirror to measure the strength of the quantum fluctuations of the vacuum. We demonstrated an \textit{in situ} modulation of the  fluctuations by a factor of 9.8 by effectively moving the atom in and out of a node of the fluctuations.  The lower limit of the strength of vacuum fluctuations we observe is 0.02 quanta, showing that we can effectively hide the atom from vacuum fluctuations. This result suggests new directions for the engineering of the vacuum.

\clearpage

\end{document}